\documentclass[a4paper,11pt,headings=big,DIV=12]{scrartcl}
\usepackage{amsmath,amssymb}
\usepackage[bookmarksdepth=2]{hyperref}
\usepackage{graphicx}
\usepackage{multirow}
\usepackage{arydshln}
\usepackage{bbm}
\usepackage{cite}
\usepackage[normalem]{ulem}
\usepackage{subfigure}
\usepackage{arydshln}

\pdfoutput=1

\addtokomafont{disposition}{\rmfamily\boldmath}
\let\tmptitle\title\renewcommand{\title}[1]{\tmptitle{\LARGE #1}}
\let\tmpauthor\author\renewcommand{\author}[1]{\tmpauthor{\large #1}}
\let\tmpdate\date\renewcommand{\date}[1]{\tmpdate{\normalsize #1}}
\newcommand{\abstrct}[1]{\begin{abstract}\vspace{-2em}\small\noindent#1\end{abstract}}

\title{
Gauged flavour symmetry for the light generations
}
\date{\vspace{-1em}}
\author{Raffaele Tito D'Agnolo$^{a,b}$\footnote{\tt\href{mailto:raffaele.dagnolo@sns.it}{raffaele.dagnolo@sns.it}}
~and David M. Straub$^{a}$\footnote{\tt\href{mailto:david.straub@sns.it}{david.straub@sns.it}}
\\ \normalsize\itshape
$^{a}$ Scuola Normale Superiore and INFN, Piazza dei Cavalieri 7, 56126 Pisa, Italy
\\ \normalsize\itshape
$^{b}$ Institute for Advanced Study, Princeton 08540, USA
}

\newcommand{\nn}{\nonumber}
\allowdisplaybreaks[2]
\newcommand{\ba}{\begin{eqnarray}}
\newcommand{\ea}{\end{eqnarray}}
\newcommand{\be}{\begin{equation}}
\newcommand{\ee}{\end{equation}}

\begin{document}

\maketitle

\abstrct{%
We study the phenomenology of a model where an $SU(2)^3$ flavour symmetry acting on the first two generation quarks is gauged and Yukawa couplings for the light generations are generated by a see-saw mechanism involving heavy fermions needed to cancel flavour-gauge anomalies.
We find that, in constrast to the $SU(3)^3$ case studied in the literature, most of the constraints related to the third generation, like electroweak precision bounds or $B$ physics observables, can be evaded, while characteristic collider signatures are predicted.
}

\section{Introduction}

Extensions of the Standard Model (SM) have to cope with the fact that the SM picture of flavour violation provides an excellent description of existing experimental results. This has motivated the idea of Minimal Flavor Violation, postulating that the SM Yukawa couplings be the only spurions breaking the global $U(3)^3$ flavour symmetry 
in the quark sector
\cite{Chivukula:1987py,Buras:2000dm,D'Ambrosio:2002ex}.
An interesting (albeit not necessary) possibility is to consider the flavour symmetry as a fundamental symmetry of nature and the spurions as background values of fundamental fields (see \cite{Feldmann:2009dc,Albrecht:2010xh,Alonso:2011yg} for recent work in this direction). Then, to avoid the appearance of Goldstone bosons, one is forced to consider {\em gauged} flavour symmetries. Recently, the observation was made that heavy fermions required to cancel the anomalies associated with a gauged $SU(3)^3$ flavour symmetry naturally lead to a see-saw mechanism for quark masses and an inverse dependence of the Yukawa couplings on the background values of the spurion fields \cite{Grinstein:2010ve}. This in turn leads to a suppression by small Yukawa couplings of flavour violating operators mediated by flavour gauge bosons (FGBs), allowing a symmetry breaking scale not far above the electroweak scale, with interesting implications for phenomenology \cite{Grinstein:2010ve,Feldmann:2010yp,Guadagnoli:2011id,Buras:2011wi}.

Here, we discuss an application of the ideas in ref.~\cite{Grinstein:2010ve}, gauging only a subgroup
\begin{equation}
G_F=SU(2)_Q \otimes SU(2)_U \otimes SU(2)_D
\end{equation}
of the $U(3)^3$ flavour group in the quark sector. Since the third generation quarks have large Yukawa couplings and are weakly mixed with the light generations, a $U(2)$ symmetry acting on the first two generations is known since a long time to provide a good starting point for flavour models \cite{Pomarol:1995xc,Barbieri:1995uv}.
Recently, a global $U(2)^3$ symmetry has been considered as an alternative to MFV \cite{Barbieri:2011ci,Barbieri:2011fc}. The most stringent constraints on the original $SU(3)^3$ flavour gauge model arise from observables related to the third generation, such as electroweak precision observables \cite{Grinstein:2010ve} or $B$ physics \cite{Buras:2011wi}. Our motivation is therefore to study whether a flavour gauge symmetry for the light generations can evade these constraints, while still giving testable predictions for flavour and collider physics.

The paper is organized as follows.
\begin{itemize}
 \item In sec.~\ref{sec:fields}, we define the field content of our model. On top of a two-family analogue of the model in \cite{Grinstein:2010ve}, we require an additional spurion to communicate between the light generations and the third one
 and $G_F$ singlet fermionic partners for the third generation quarks.
 \item In sec.~\ref{sec:spec}, we discuss the spectrum of the 9 flavour gauge bosons, the 6 fermionic partners of the quarks and some aspects of the scalar excitations of the $G_F$ breaking fields.
 \item In sec.~\ref{sec:flavour}, we derive the bounds from flavour physics on the model. In particular, tree-level mediated contributions to $\Delta S=2$ set strong constraints on some of the FGB masses. $B$ physics is unaffected and the second generation fermionic partners can still be light.
 \item In sec.~\ref{sec:collider}, we present the LHC phenomenology of the model, given the flavour constraints. We show that interesting effects related to the second generation fermionic partners are possible. Not involving third generation quarks, these signatures are markedly different from the $SU(3)^3$ case.
\end{itemize}

\section{Field content}\label{sec:fields}

Gauging the flavour $SU(2)^3$ gives rise to anomalies which can be cancelled by the addition of heavy chiral fermions. Although the cubic $SU(2)$ anomalies automatically cancel in constrast to the $SU(3)$ ones, the field content required to cancel the mixed SM-flavour anomalies is analogous in the $SU(2)$ and $SU(3)$ cases.
For the first two generation quark fields and their heavy partners, we thus choose the field content shown in the first column of table~\ref{tab:fields}, following the notation in \cite{Grinstein:2010ve}. The quark fields of the first two generations are accompanied by $SU(2)_L$ singlet flavoured heavy quark fields and by the two scalars $S_u$ and $S_d$ that transform as bidoublets under $G_F$.

\begin{table}[tbp]
\renewcommand{\arraystretch}{1.5}
\begin{center}
\begin{tabular}{lccc}
\hline
&1st/2nd gen. fermions&3rd gen. fermions& scalars\\
\hline
\multirow{3}{*}{SM fields}&$Q_L \sim (2,1,1)_{Q_L}$ &$q_L \sim (1,1,1)_{Q_L}$ &$H \sim (1,1,1)_{H}$\\
&$U_R^c \sim (1,2,1)_{U_R^c}$ &  $T_R^c \sim (1,1,1)_{U_R^c}$\\
&$D_R^c \sim (1,1,2)_{D_R^c}$ & $B_R^c \sim (1,1,1)_{D_R^c}$\\
\hline
\multirow{4}{*}{new fields}&$\psi_{u} \sim (1,2,1)_{U_R}$ &  {$\psi_{q} \sim (1,1,1)_{Q_L}$}
&$S_u \sim (2,2,1)_{1}$\\
&$\psi_{u_R}^c \sim (2,1,1)_{U_R^c}$ & {$\psi_{q_R}^c \sim (1,1,1)_{Q_L^c}$}
&$S_d \sim (2,1,2)_{1}$\\
&$\psi_{d} \sim (1,1,2)_{D_R}$ & 
&  $V \sim (2,1,1)_{1}$\\
&$\psi_{d_R}^c \sim (2,1,1)_{D_R^c}$ &\\
\hline
\end{tabular}
\end{center}
\renewcommand{\arraystretch}{1.0}
\caption{Field content of the model. The numbers in brackets indicate the representation under $G_F$, while the subscript refers to the transformation under $G_\text{SM}$.
}
\label{tab:fields}
\end{table}

In our setup, the third generation quark fields are $G_F$ singlets. To introduce a communication between the third and the light generations, two ingredients are needed: a scalar field being a doublet under one of the flavour gauge group factors and additional heavy fermion fields which are $G_F$ singlets and vector-like under the SM gauge group $G_\text{SM}=SU(3)_c\otimes SU(2)_L \otimes U(1)_Y$, so as not to spoil the SM anomaly cancellation.
We choose the additional fermions to be $SU(2)_L$ doublets and the additional scalar $V$  to be a doublet under $SU(2)_Q$. The resulting field content is shown in the second column of table~\ref{tab:fields}. We have also examined different possibilities, that include taking as third generation partners $SU(2)_L$ singlets and adding different spurions to link the third generation to the light ones. We have found that in general FCNC constraints are more severe in these cases, placing the new particles well outside the reach of present experiments.

The most general renormalizable Lagrangian\footnote{We have omitted a term of the form $\bar q_L \psi_{q_R}$,
which can be removed by a redefinition of the fields $q_L$ and $\psi_q$.} involving these fields reads
\begin{equation}
\mathcal L = \mathcal L_\text{kin} - V(H,V,S_u,S_d)
+ \mathcal L_{ll} + \mathcal L_{hh} + \mathcal L_{lh} \,,
\end{equation}
where
\begin{align}
\mathcal L_{ll} & = - y_t \bar q_L \tilde H T_R - y_b \bar q_L H B_R + \text{h.c.}\,,
\\
\mathcal L_{hh} & =
\lambda_u' \bar\psi_u S_u \psi_{u_R} + \lambda_d' \bar\psi_d S_d \psi_{d_R}  + M_{q}' \bar \psi_q \psi_{q_R} +\text{h.c.}\,,
\\
\mathcal L_{lh} & = 
\lambda_u \bar Q_L \tilde H \psi_{u_R} + M_u \bar\psi_u U_R
+
\lambda_d \bar Q_L H \psi_{d_R} + M_d \bar\psi_d D_R
+ \lambda_q \bar Q_L V \psi_{q_R}
\nn\\&
+ \lambda_t \bar \psi_q \tilde H T_R
+ \lambda_b \bar \psi_q H B_R
+\text{h.c.}\,.
\label{eq:L}
\end{align}

We assume an approximate discrete symmetry distinguishing the new fermion fields from the SM fermions. Consequently, the dimensionless parameters in $\mathcal L_{hh}$ and $\mathcal L_{ll}$ are assumed to be of order~1, while those in $\mathcal L_{lh}$ are assumed to be small (in practice, of order $10^{-2}$ or smaller).

After integrating out the heavy fermions, the Lagrangian in (\ref{eq:L}) leads to the effective Yukawa Lagrangian (using the same symbol to denote scalar fields and their VEVs)
\begin{equation}
 \begin{split}
\mathcal L_Y^\text{eff} =
- y_t\; \bar q_L \tilde H \,T_R
- y_b\; \bar q_L H \,B_R
- \bar Q_L \tilde H \left( \frac{\lambda_t \lambda_q  V}{M_q'} \right) T_R
- \bar Q_L  H \left( \frac{\lambda_b \lambda_q  V}{M_q'} \right) B_R
\\
- \bar Q_L \tilde H \left( \frac{\lambda_u M_u}{\lambda_u' S_u} \right) U_R
- \bar Q_L H \left( \frac{\lambda_d M_d}{\lambda_d' S_d} \right) D_R
\,.
\label{eq:Leff}
 \end{split}
\end{equation}
Communication between the third generation and the first two only occurs between right-handed 3rd generation fields and left-handed light generation fields. This is due to the choice of a single spurion transforming as $(2,1,1)$ under $G_F$ and reproduces the Yukawa structure found in \cite{Barbieri:2011ci} for a 
global $U(2)^3$ with an analogous breaking pattern.
The smallness of the first and second generation Yukawa couplings is due to the smallness of the parameters $\lambda_{u,d,t,q}$ with respect to $\lambda_{u,d}'$. 
A priori, we do not assume any hierarchy\footnote{The hierarchy among the {\em first two} generation masses arises from a hierarchy in the VEVs of $S_{u,d}$ and is unexplained also in our framework.} between $S_{u,d}$ and $M_{u,d}$ or between $V$ and $M_q'$.

We also note that with the above scalar field content, it is possible to construct a renormalizable potential giving rise to the Yukawa pattern of the SM. This is an interesting observation since in the $SU(3)^3$ case, this was found not to be possible~\cite{Grinstein:2010ve}.

\section{Spectrum}\label{sec:spec}

\subsection{Spin 1}

The masses of the flavour gauge bosons (FGBs) arise from the terms
\begin{align}
\mathcal L_\text{mass} &=
 \text{tr}\left|g_Q A_Q S_u + g_U S_u A_U\right|^2
+\text{tr}\left|g_Q A_Q S_d + g_D S_d A_D\right|^2
+\text{tr}\left|g_Q A_Q V \right|^2
\,,
\end{align}
where $A_{Q}=A_{Q}^a\sigma^a/2$ etc. A numerical diagonalization of the mass matrix is straightforward, however 
we will see in sec.~\ref{sec:fgbfcnc} that flavour constraints require a hierarchy  $V\gg S_{u,d}$.
In this limit, three FGBs, which are dominantly the ones associated to $SU(2)_Q$, will be heavy and have almost the same mass $M_{A_Q}\approx g_QV/\sqrt2$.
The six remaining bosons form two sets of three nearly degenerate states each, with masses given approximately by (up to small mixings of order $y_u/y_c$ and $y_d/y_s$)
\begin{align}
M_{A_U}^2 &= g_U^2 \left(\frac{\lambda_u}{\lambda_u' y_u}\right)^2 M_u^2
\,,\label{eq:MAU2}\\
M_{A_D}^2 &= g_D^2 \left(\frac{\lambda_d}{\lambda_d' y_d}\right)^2 M_d^2
\,.\label{eq:MAD2}
\end{align}
In the opposite limit $S_{u,d}\gg V$, on the other hand, there is always one light FGB with mass of order $g_QV$.
The reason is that, in the absence of $V$, there is a residual $U(1)$ unbroken by $S_u$ and $S_d$.

\subsection{Spin \texorpdfstring{$\tfrac{1}{2}$}{1/2}}\label{F_mass}

The $6\times6$ mass matrices for up- and down-type fields can be read off eq.~(\ref{eq:L}).
Their numerical diagonalization can be easily performed and is discussed in appendix~\ref{sec:fermiondiag}. Here, we present approximate analytical expressions for them. We first observe that the VEVs of $S_{u,d}$ can be brought to the following form by $SU(2)$ rotations,
\begin{equation}
\left(\frac{\lambda_{u,d}}{\lambda_{u,d}'}M_{u,d}\right) S_{u,d}^{-1} =  U_{Q_{u,d}}^T
\begin{pmatrix}
y_{u,d}&0\\
0&y_{c,s}
\end{pmatrix}
,
\end{equation}
where $U_{Q_{u,d}}$ are $2\times2$ rotation matrices that can be written as (cf. the appendix of \cite{Barbieri:2011ci})
\begin{equation}
U_{Q_u} = \begin{pmatrix} c_ue^{i\phi_u} & s_u e^{i\alpha_u} \\ -s_ue^{-i\alpha_u} & c_ue^{-i\phi_u} \end{pmatrix},
\qquad
U_{Q_d} = \begin{pmatrix} c_de^{i\phi_d} & s_de^{i\alpha_d} \\ -s_de^{-i\alpha_d} & c_de^{-i\phi_d} \end{pmatrix}.
\label{eq:UQ}
\end{equation}
Then, one can perform the following rotations on second and third generation quark fields and their partners,
\begin{align}
\begin{pmatrix}
U_L^2 \\ T_L
\end{pmatrix}
&\to
\begin{pmatrix}
c_t & s_t \\
-s_t & c_t
\end{pmatrix}
\begin{pmatrix}
U_L^2 \\ T_L
\end{pmatrix}
\,,\qquad
s_t/c_t=\frac{\lambda_q\lambda_t V}{y_t M_q'}
\,,
\label{eq:rotT}
\\
\begin{pmatrix}
U_L^2 \\ \psi_t
\end{pmatrix}
&\to
\begin{pmatrix}
c_q & s_q \\
-s_q & c_q
\end{pmatrix}
\begin{pmatrix}
U_L^2 \\ \psi_t
\end{pmatrix}
\,,\qquad
s_q/c_q=\frac{\lambda_q V}{M_q'}
\,,
\label{eq:rotq}
\end{align}
followed by the redefinitions
\be
U_L \to U_{Q_u}^* U_L, \quad \psi_{u_R} \to U_{Q_u}^* \psi_{u_R}\, .
\label{eq:rotU}
\ee
At this point the $6\times 6$ mass matrix is approximately\footnote{%
Up to small terms of  order $\epsilon^3$, where $\epsilon=\lambda_{u,t,q}$ or $s_u$. Although the rotation (\ref{eq:rotT}) is not strictly necessary at $O(\epsilon^2)$, we show it because it generates 23-mixing in the CKM matrix. In our numerical analysis, we work with the exact expressions.
}
block diagonal and we can diagonalize it with $2\times 2$ rotations, that we write as
\begin{align}
\begin{pmatrix}
u_{L(R)} \\ u'_{L(R)}
\end{pmatrix}
&=
\begin{pmatrix}
c_{L(R)}^u & -s_{L(R)}^u \\
s_{L(R)}^u& c_{L(R)}^u
\end{pmatrix}
\begin{pmatrix}
U_{L(R)}^1 \\ \psi_{u_{(R)}}^1
\end{pmatrix}
\,,&
\begin{pmatrix}
c_{L(R)} \\ c'_{L(R)}
\end{pmatrix}
&=
\begin{pmatrix}
c_{L(R)}^c & -s_{L(R)}^c \\
s_{L(R)}^c& c_{L(R)}^c
\end{pmatrix}
\begin{pmatrix}
U_{L(R)}^2 \\ \psi_{u_{(R)}}^2
\end{pmatrix}
\,,\\
\begin{pmatrix}
t_{L(R)} \\ t'_{L(R)}
\end{pmatrix}
&=
\begin{pmatrix}
c_{L(R)}^t & -s_{L(R)}^t \\
s_{L(R)}^t& c_{L(R)}^t
\end{pmatrix}
\begin{pmatrix}
T_{L(R)} \\ \psi_{q_{(R)}}
\end{pmatrix}
\,,
\label{eq:sLR}
\end{align}
and analogously for down-type fields.

To an excellent approximation\footnote{That is keeping only the leading order in $\left(M_{u,d}/m_{q'}\right)$ and $\left(\lambda v/m_{q'}\right)$, where $\lambda=\lambda_{u,d, b,t, q}\,$.}%
, this leads to the following masses for the 6 quarks and their heavy partners,
\begin{align}
m_q &= v y_q
\,,&
m_{u,c}' &\approx \frac{\lambda_u}{y_{u,c}} M_u
\,,&
m_{d,s}' &\approx \frac{\lambda_d}{y_{d,s}} M_d
\,,&
m_{t,b}' &\approx M_q'
\,.
\label{eq:mq}
\end{align}
Consequently, the first generation heavy quarks are heavier than the second generation ones by roughly a factor of 400 for up-type quarks and 20 for down-type quarks, while the partners of the top and bottom are degenerate but a priori unrelated to the first two generation ones.

Following the same procedure we obtain also the mixings between the new fermions and the SM fermions
\begin{align}
s_L^{u} &\approx  \frac{m_u}{M_u}
\,,&
s_L^{c} &\approx  \frac{m_c}{M_u}
\,,&
s_L^{t} &\approx \frac{\lambda_t v}{M_q'}\frac{m_t}{M_q'}
\label{eq:sL}
\,,\\
s_R^{u} &\approx  \frac{y_u}{\lambda_u}
\,,&
s_R^{c} &\approx  \frac{y_c}{\lambda_u}
\,,&
s_R^{t} &\approx \frac{\lambda_tv}{M_q'}
\,,
\label{eq:sR}
\end{align}
and analogously for the down-type sector. Combining eqs. (\ref{eq:mq})--(\ref{eq:sR}), we obtain the approximate relations
\begin{align}
s_L^{u}s_R^u &\approx  \frac{m_u}{m_u'}
\,,&
s_L^{c}s_R^c &\approx  \frac{m_c}{m_c'}
\,,&
s_R^{t} &\approx  \frac{m_t}{m_t'}\,\frac{\lambda_t}{y_t}
\,,&
s_L^{t} &\approx  \frac{m_t}{m_t'}\,s_R^t
\,.
\label{eq:sRapp}
\end{align}
All these mixing angles are small, since, as mentioned in section~\ref{sec:fields}, we assume $\lambda_{u,d}$ and $\lambda_{t,b}$
to be small.
The possible exception is the right-handed mixing of the second generation.
Indeed, for our approximations to be valid, we will require $\lambda_u>y_c$, otherwise $\psi_{u_R}^2$ and $U_R^2$ exchange roles as dominant components of $c_R$ and $c'_R$. In the down-type sector, we analogously require $\lambda_d>y_s$. For the third generation, there is a double suppression in the right-handed mixing since $m_{t'}\gg m_t$ and $\lambda_t\ll y_t$ and a triple suppression in the left-handed mixing.

If we neglect terms of $\mathcal{O}(s_L^t s_L^b)$, the rotations (\ref{eq:rotT})--(\ref{eq:sLR}) lead to the following form for the effective CKM matrix $V$,
\begin{equation}
V_{ij} = c_{L}^{u^i}c_{L}^{d^j}(U_{u_L}^* \, U_{d_L}^T)_{ij}
\label{eq:ckm}
\end{equation}
where there is no sum implied over $i$, $j$ and
\begin{equation}
U_{u_L}=
\begin{pmatrix}
\multicolumn{2}{c}{\multirow{2}{*}{$U_{Q_{u_L}}$}} & 0 \\
&&0\\
0 & 0 & 1
\end{pmatrix}
\begin{pmatrix}
1&0&0\\
0&c_t&s_t\\
0&-s_t&c_t
\end{pmatrix}
\end{equation}
and analogously for $U_{d_L}$. The corrections to (\ref{eq:ckm}) are non-unitary due to the mixing between light and heavy states.
Since the largest left-handed mixings occur among the second generation fermions, only the second row and column of the CKM matrix could be appreciably affected. Given the direct measurements of CKM elements not assuming unitarity \cite{Nakamura:2010zzi}, the strongest constraints then come from the first two rows of the CKM matrix.
First row unitarity leads to $|s_L^{s}|\lesssim0.21$ at $2\sigma$; since
$s_L^{s}\approx \lambda_d v/ m'_{s} $ and $\lambda_d$ is small by assumption, this bound is trivially fulfilled. Second row unitarity leads to the bound $(s_L^{s})^2+(s_L^{c})^2<\left(0.22\right)^2$ at $2\sigma$, which is comparably weak.

A fit of the relevant parameters in (\ref{eq:UQ}) and (\ref{eq:rotT}) to tree-level CKM constraints, i.e. $|V_{ud}|$, $|V_{us}|$, $|V_{ub}|$, $|V_{cb}|$, $\sin(2\beta)$ and $\gamma$, gives the following results,
\begin{align}
s_u &= 0.086\pm0.003
\,,\\
s_d &= -0.22\pm0.01
\,,\\
\alpha_u-\alpha_d &= (97\pm 9)^\circ
\,,\\
s_b-s_t=
\frac{\lambda_qV}{M_{q'}}
\left(
\frac{\lambda_b}{y_b}
-
\frac{\lambda_t}{y_t}
\right)
&= 0.0411 \pm 0.0005 = |V_{cb}|
\label{eq:Vcb}
\,.
\end{align}

\subsection{Spin 0}

Of the $8+8+4$ real scalar fields contained in $S_u$, $S_d$ and $V$, $9$ get ``eaten'' by the flavour gauge bosons, leaving 11 physical flavoured SM singlet scalar fields or ``flavons''. They correspond to fluctuations of first two generation quark masses, the angles and phases contained in $U_{Q_{u,d}}$ and the VEV $V$.

While the details of the flavon spectrum depend on the scalar potential, which we do not specify, we will discuss some aspects of the flavour-diagonal flavons, i.e. the fields corresponding to fluctuations of the light quark masses, in section~\ref{sec:collider}. The lighter of the diagonal modes in $S_u$ and $S_d$, which we will call $S_c$ and $S_s$, are assumed to have 
masses similar to the corresponding VEVs,
\begin{align}
M_{S_c} &\sim \langle S_c \rangle = \frac{\lambda_u M_u}{\lambda_u'y_c} = \frac{m_c'}{\lambda_u'}
\,,&
M_{S_s} &\sim \langle S_s \rangle = \frac{\lambda_d M_d}{\lambda_d'y_s} = \frac{m_s'}{\lambda_d'}
\,,
\label{eq:MS}
\end{align}
so according to our assumption that $\lambda_{u,d}'$ be of order~1, we expect the masses of these flavons to be comparable to those of the second generation heavy fermions.
Analogously to the discussion in appendix~A.2 of ref.~\cite{Grinstein:2010ve}, one can show that
$S_{c,s}$ couple to second generation quarks approximately as
\begin{equation}
-\lambda_u' s_L^cs_R^c \, \bar c_L c_R S_c
-\lambda_d' s_L^ss_R^s \, \bar s_L s_R S_s
\approx
-\frac{m_c}{\langle S_c \rangle} \bar c_L c_R S_c
-\frac{m_s}{\langle S_s \rangle} \bar s_L s_R S_s
\,.
\end{equation}
Similarly, the flavour off-diagonal modes can be shown to have even more strongly suppressed couplings to SM fermions.

\section{Bounds from flavour physics}\label{sec:flavour}

\subsection{Spin 1}\label{sec:fgbfcnc}

The FGBs can mediate $\Delta F=2$ transitions at tree level. Integrating them out, one obtains dimension 6 four-quark operators which read for down-type quarks
\begin{align}
\label{eq_Qi}
Q_1^{ji} &= (\bar d_i^\alpha \gamma^\mu P_L d_j^\alpha)(\bar d_i^\beta \gamma_{\mu} P_L d_j^\beta)
,\nn &
\widetilde Q_1^{ji} &= (\bar d_i^\alpha \gamma^\mu P_R d_j^\alpha)(\bar d_i^\beta \gamma_{\mu} P_R d_j^\beta)
,\nn &
Q_5^{ji} &= (\bar d_i^\alpha P_L d_j^\beta)(\bar d_i^\beta P_R d_j^\alpha)
.
\end{align}
In the quark mass eigenstate basis, the Wilson coefficients of the above operators are approximately given by (cf.~\cite{Guadagnoli:2011id,Buras:2011wi})
\begin{align}
C_1^{ji} &= - \frac{g_Q^2}{2} (M^2_V)^{-1}_{a,b} (U_{dL}^{l\dagger}  \, T^a \,  U_{dL}^l)_{ij}
(U_{dL}^{l\dagger}  \, T^b \,  U_{dL}^l)_{ij}
~,\nn \\
\widetilde C_1^{ji} &= - \frac{g_D^2}{2} (M^2_V)^{-1}_{6+a,6+b} (U_{dR}^{l\dagger}  \, T^a \,  U_{dR}^l)_{ij}
(U_{dR}^{l\dagger}  \, T^b \,  U_{dR}^l)_{ij}
~,\nn \\
C_5^{ji} &= 2g_Q g_D (M^2_V)^{-1}_{a,6+b} (U_{dL}^{l\dagger}  \, T^a \,  U_{dL}^l)_{ij}
(U_{dR}^{l\dagger}  \, T^b \,  U_{dR}^l)_{ij}
~,
\end{align}
where a sum over $a$ and $b$ in the range $1,2,3$ is understood and
\begin{equation}
T^a  =
\begin{pmatrix}
\sigma^a/2 & 0 \\
0 & 0
\end{pmatrix}.
\label{eq:gen}
\end{equation}
The fermion mixing matrices are defined in appendix~\ref{sec:fermiondiag}.
Eq.~(\ref{eq:gen}) reminds us that the third generation quarks and their partners only couple to the FGBs via mixing with the first two generation fermions. Consequently, the strongest constraint on the FGB masses comes from the kaon system, while contributions to $B_d$ and $B_s$ mixing are suppressed.

The most constraining operator is the left-right operator $Q_5^{12}$, which is enhanced by renormalization group running and a chiral factor. The dominant contribution reads approximately (in the standard CKM phase convention)
\begin{equation}
C_5^{12} \approx \frac{4s_d^2}{V^2} ~ e^{-2i\beta_\text{CKM}}\,.
\end{equation}
%
The experimental bound on $C_5^{12}$ \cite{Bona:2007vi} thus requires
\begin{equation}
V\gtrsim 10^8 \text{ GeV} \,,
\label{eq:Vbound}
\end{equation}
with weak dependence on other parameters. Once this bound is satisfied, the remaining FGB masses are essentially unconstrained.
We also checked that in this case, the contributions of $\Delta S=1$ operators to FCNCs or EDMs as well as contributions to $\Delta C=1,2$ observables are completely negligible.

\subsection{Spin \texorpdfstring{$\tfrac{1}{2}$}{1/2}}\label{sec:DF2}

Given the strong bound in (\ref{eq:Vbound}), one should expect that also the partners of third generation quarks have masses of that order, unless one introduces an {\em ad hoc} hierarchy $M_{q'}\ll V$ and uses a tiny coupling $\lambda_q$ to fulfill condition~(\ref{eq:Vcb}). We will not consider this possibility in the following and assume the $t'$ and $b'$ to be decoupled from low energy phenomenology. This choice renders corrections to the $Zb\bar b$ vertex completely negligible. Since in addition the $t'$ and $b'$ are nearly degenerate the oblique electroweak parameters are also almost unaffected. For example, the 3rd generation contribution to the $T$ parameter reads
\begin{equation}
\Delta T^{t',b'} \propto \left((s_R^t)^2-(s_R^b)^2\right)^2 \frac{M_{q}^{\prime 2}}{v^2} \approx \left((\lambda_t)^2-(\lambda_b)^2\right)^2\frac{v^2}{M_{q}^{\prime 2}}
\,.
\end{equation}
Similarly, $\Delta S$ becomes zero in the limit of large $M_q '$, when we are left with the SM and a decoupled vector-like doublet. It is easy to verify numerically that contributions from second generation partners are equally unimportant. For analytical expressions we refer to the appendix of~\cite{Grinstein:2010ve}, whose results apply to our first two generations. 

The relevant effects of the new fermions on flavour violating observables will come from the second generation partners. Concretely, the $c'$ enters in box diagram contributions to $K^0$-$\bar K^0$ mixing. $B_d$ and $B_s$ mixing, on the other hand, are virtually unaffected.

\begin{figure}[tbp]
\centering%
\includegraphics[width=0.5\textwidth]{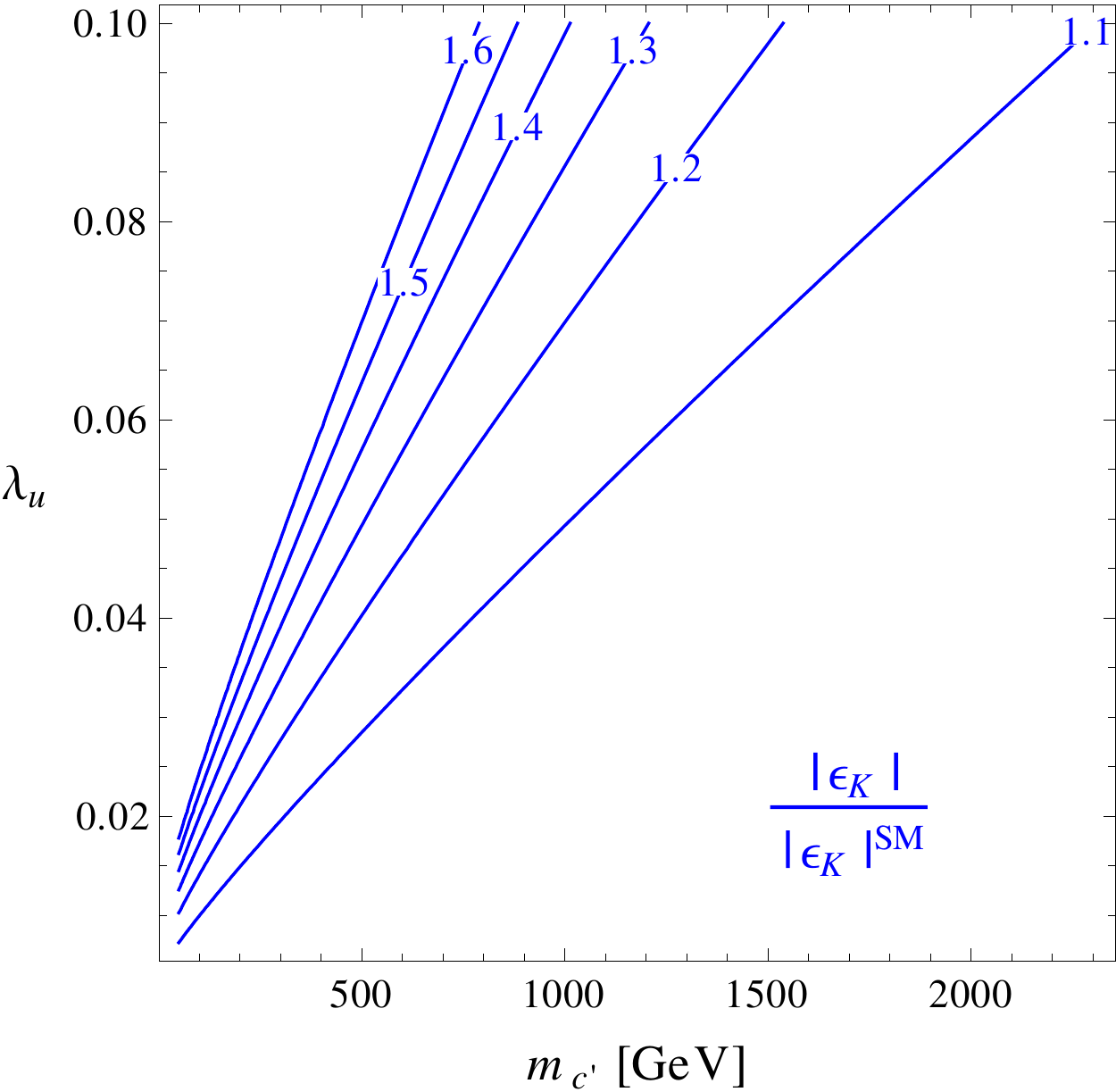}
\caption{Ratio of the model prediction for $|\epsilon_K|$ over the SM expectation in terms of $\lambda_u$ and the $c'$ mass. The dependence on other model parameters is negligible.}
\label{fig:epsK}
\end{figure}

The $\Delta F=2$ box contributions have recently been considered in~\cite{Buras:2011wi} in the context of the original $SU(3)^3$ flavour gauge model. Our analysis proceeds analogously, except that the new physics contributions are now dominated by $c'$ contributions instead of $t'$. As a result, the heavy fermions lead to a modification of the charm-top contribution to the $\Delta S=2$ amplitude  (and, to a lesser extent, the charm contribution). With $s_L^{u,t}\ll s_L^c$ and $m_{u',t'}\gg m_{c'}$, the result depends only on $s_L^c$ and $m_{c'}$. Using additionally $s_L^{c}\approx \lambda_u v/ m'_{c} $, fig.~\ref{fig:epsK} shows the impact on the imaginary part of the $\Delta S=2$ amplitude in the $m_{c'}$-$\lambda_u$ plane. We make two observations:
\begin{itemize}
\item As in the $SU(3)^3$ case, we find that the box contributions always increase $|\epsilon_K|$, which is interesting in view of the discrepancy between the measured value $(2.229\pm0.010)\times10^{-3}$ and the SM expectation, $(1.82\pm0.28)\times10^{-3}$ \cite{Brod:2011ty}.
\item Given the above discrepancy, the 22\% uncertainty of the SM result and the larger uncertainties in the QCD corrections of the top-charm and charm contributions to $\epsilon_K$ compared to the top contribution \cite{Brod:2010mj,Brod:2011ty}, a value of $|\epsilon_K/\epsilon_K^\text{SM}|=1.6$ (for central values of the parameters) cannot be excluded. Since we assume the heavy-light coupling $\lambda_u$ to be small, we see from fig.~\ref{fig:epsK} that the box contributions generically do not lead to excessive $|\epsilon_K|$ even for light $m_{c'}$.
\item The corresponding $c'$ box contributions to $B_d$ and $B_s$ mixing are negligible: if $|\epsilon_K/\epsilon_K^\text{SM}|<1.6$, their SM amplitudes are modified by less than 1\%.
\end{itemize}

The $B\to X_s\gamma$ decay is not a relevant constraint in our setup. With the $t'$ heavy and flavour gauge contributions being completely negligible, the dominant contribution stems from a diagram with internal $c'$ exchange. Denoting the SM loop function as $f(x_t)$, where $x_i=m_i^2/m_W^2$, the contribution is obtained by replacing $f(x_t)$ with $-(s_L^c)^2f(x_{c'})$, which is always negligible even for extreme values of the parameters.

\section{Collider signatures}\label{sec:collider}

\begin{figure}[tbp]
\centering%
\includegraphics[width=0.6\textwidth]{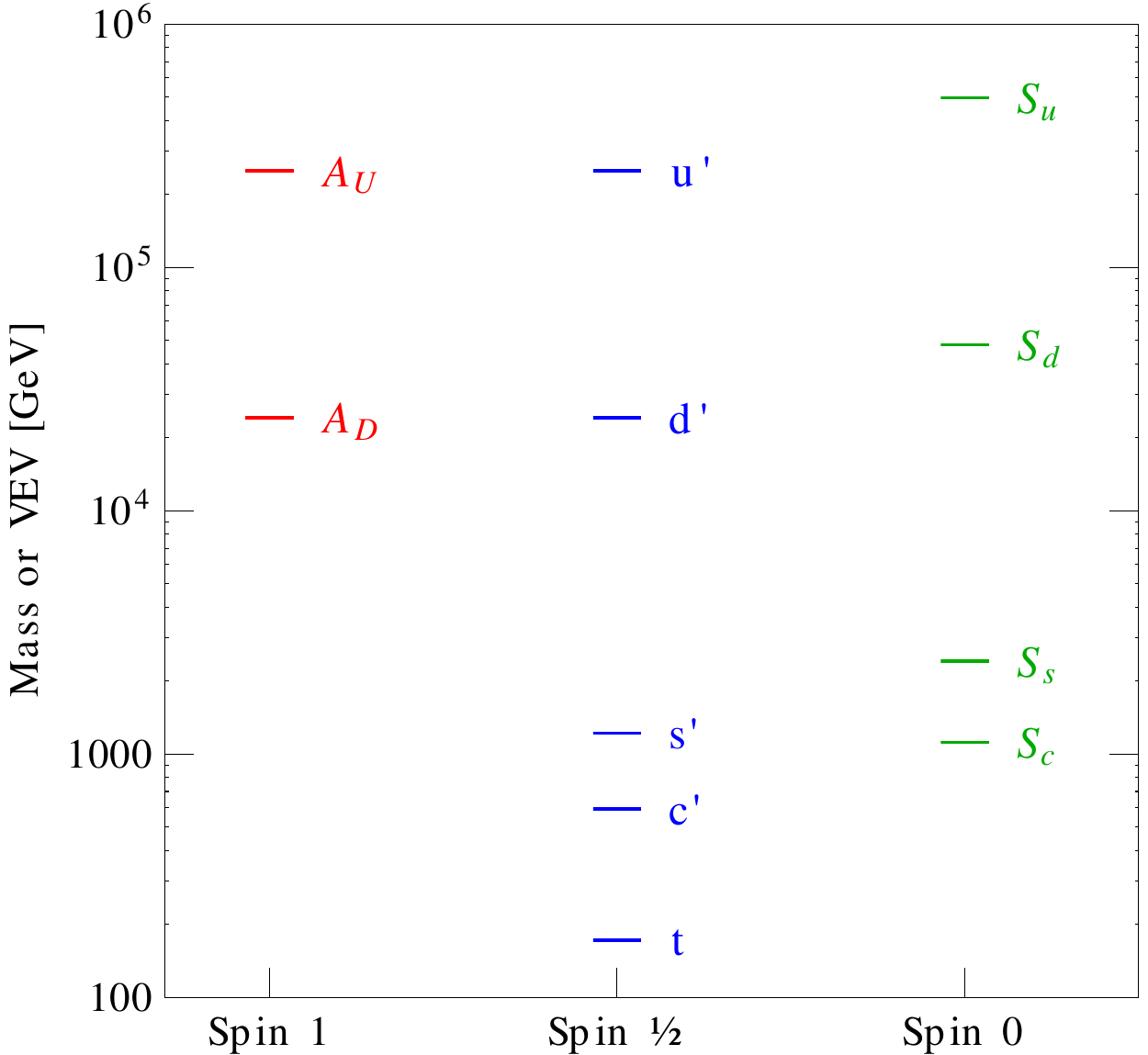}
 \caption{Typical spectrum. The FGBs of $SU(2)_Q$ , the VEV $V$ as well as the $t'$ and $b'$ are heavier than $10^8$~GeV and are not shown. In the scalar sector, only the VEVs of the flavour-diagonal flavons are shown.}
\label{fig:spec}
\end{figure}

Before discussing the collider signatures of the model, we summarize the features of the spectrum, given the bounds from flavour physics found in the previous section.

\begin{itemize}
\item The FGB associated to $V$ has to be heavier than roughly $\sqrt{2}/g_Q\times10^8$\,GeV. The same is true for the third generation partners, unless one tunes $\lambda_q$, as discussed at the beginning of section~\ref{sec:DF2}.
\item The masses of second generation fermion partners are virtually unconstrained by flavour physics, as long as $\lambda_u$ is not too large, as shown in fig.~\ref{fig:epsK}.
\item The partners of first generation quarks are always heavier than second generation ones by a factor $m_{c,s}/m_{u,d}$. The same is true for the VEVs of the flavons $S_{u,d}$ and, assuming $g_{U,D}/\lambda_{u,d}'=O(1)$ for all the remaining FGBs.
\end{itemize}
Consequently, the only states which are accessible to direct production are the second generation fermion partners $s'$ and $c'$ as well as the flavour diagonal flavons $S_s$ and $S_c$, whose production and decay we will now discuss.

Fig.~\ref{fig:spec} shows a typical spectrum, obtained with the following set of parameters:

\begin{center}
\renewcommand{\arraystretch}{1.5}
\begin{tabular}{cccccccccc}
$g_{Q,U,D}$ & $\lambda_{u,d}'$ & $\lambda_u$ & $\lambda_d$ & $\lambda_t$ & $\lambda_b$ & $\lambda_q$ &
$M_{u,d}$ & $M_{q}'$ & $V$
\\\hline
0.5 & 0.5 & 0.01 & 0.002 & 0.001 & 0.01 & 0.068 & 200 GeV & $10^8$ GeV & $10^8$ GeV
\end{tabular}
\renewcommand{\arraystretch}{1.0}
\end{center}

\subsection{Spin \texorpdfstring{$\tfrac{1}{2}$}{1/2}}

The production mechanism of the $c'$ and $s'$ is analogous to that of fourth generation quarks \cite{Cacciari:2008zb}, while the relevant decay modes involve electroweak gauge bosons, the Higgs boson and the scalar flavons.

The most interesting channel for discovery lies in pair production, that can be followed by double Higgs production, a
$W$ or a $Z$ decay.
The flavons might or might not be relevant for these decay chains. In the following we will study the decay of the heavy $c'$, but the discussion holds for the $s'$  as well, with the obvious replacements $c\to s, u\to d$ etc.
Neglecting the light quark masses,
we can write the two body decay widths as \cite{Deshpande:1981zq, Eilam:1990zc, Cacciapaglia:2010vn}
\begin{align}
\Gamma(Q_j \to V q_i)&=  \left(|g_L^V|^2+|g_R^V|^2\right)
\frac{\left(M^2-M_V^2\right)^2 \left(M^2+2 M_V^2\right)}{32 \pi  M^3 M_V^2}
\\
\Gamma(Q_j \to S q_i)&= \left(|g_L^S|^2+|g_R^S|^2\right)
\frac{\left(M^2-M_S^2\right)^2 }{32 \pi  M^3}
\end{align}
where $V=W, Z$; $S=h, S_{s,c}$ and we have rewritten the vector and scalar couplings of the fermions as
\ba
\bar{q}_i\gamma_\mu\left((g_L^V)_{ij}P_L+(g_R^V)_{ij}P_R\right)Q_j V^\mu\,, \quad \bar{q}_i\left((g_L^S)_{ij}P_L+(g_R^S)_{ij}P_R\right)Q_j S\,.
\ea 
Clearly the constants $g_{L, R}$ set the relative importance of the different decay modes. To have an intuition about their dependence on the model parameters we can use the approximations introduced in section \ref{F_mass}.
In this limit we have, for second generations up-type quarks
\begin{align}
\left(g_L^Z\right)_{cc'} & \approx \frac{g}{2 c_W}s_{L}^{c}
\,,&
\left(g_L^W\right)_{sc'} & \approx \frac{g}{\sqrt{2}}s_{L}^{c}
\,,&
\left(g_R^h\right)_{cc'} & \approx \lambda_u
\,,&
\left(g_{L,R}^S\right)_{cc'} & \approx -\lambda_u' s_{L,R}^{c}
\,.&
\end{align}
Notice that $g_R^{W,Z}=0$ for vertices with one light and one heavy fermion of the first two generations, since the heavy fields are $SU(2)_L$ singlets. The $g_L^h$ coupling is suppressed by $s_L^cs_R^c$ and thus negligible.
Since $s_L^{c}\approx \lambda_u v/ m'_{c} $, the branching ratios to $W, \, Z$ and $h$ turn out to all be proportional to the parameter combination $\lambda_u^2m_c'$ in the limit where the $c'$ is much heavier than $W$, $Z$ and Higgs.  Then, we always have
\be
\text{BR}(c' \to Ws) \approx 2 \, \text{BR}(c' \to Zc) \approx \text{BR}(c' \to hc) 
\label{eq:BRSM}
\ee
independently of the other model parameters.
Kinematically, we now have to distinguish between the case where the flavon $S_c$ is heavier or lighter than the $c'$. According to eq.~(\ref{eq:MS}), both cases are possible, as we take $\lambda_{u,d}'$ to be of order 1 and the flavon mass to be of the order of its VEV.

\paragraph{$S_c$ heavier than $c'$}

In this case, the $W, Z$ and $h$ decays are the only modes and the branching ratios are roughly 40\%, 20\% and 40\% according to equation (\ref{eq:BRSM}).
Therefore we can expect a significant double Higgs production at the LHC
\begin{equation}
pp\to q'q'\to h h q q \,.
\end{equation}
These decays can lead to striking final states with six jets,  four of which can be b-jets or even to the production of four $W$ bosons.

While there are not yet any stringent limits on the final states with Higgses or $Z$s, the searches for decays with $W$s in the final state can be used to set bounds on the $c'$ and $s'$ masses.
Searching for heavy quarks in $WW+2$~jet final states and assuming $\text{BR}(q' \to Wq)=1$, CDF sets a bound $m_{q'} > 335$~GeV \cite{CDFtprime} and, very recently, ATLAS finds $m_{q'} > 350$~GeV at $95\%$~C.L. \cite{:2012bt}
If we assume that the acceptance times efficiency of the cuts of the latter two analyses are the same for our signal as for those used by CDF and ATLAS, then we can rescale their results with our $c'$ and $s'$ branching ratios and obtain $m_{c',s'} > 295$~GeV at $95\%$~C.L. from the CDF analysis and
a  comparable limit
from the ATLAS search.

In both cases the limit on the cross section is computed through a maximum likelihood fit of the reconstructed mass of the heavy quarks. Therefore we are assuming that the product of efficiency of the cuts and acceptance is the same bin by bin and that the mass reconstruction algorithm does not perform differently on our signal. This is not unreasonable since in all cases heavy quarks with the same quantum numbers under $SU(3)\times SU(2)\times U(1)$ have been considered, nonetheless the bound that we have obtained should be taken with some caution.

On the other hand, the indirect limits from the measurement of the Higgs production cross section by CMS and ATLAS \cite{Chatrchyan:2011tz, Aad:2011qi}, do not apply to us since the vertex $h q'q'$ always involves two mixings.

\paragraph{$S_c$ lighter than $c'$}

In this region, the new decay mode $c'\to S_c c$ opens up. Whether it dominates depends on the model parameters. In the limit of heavy $c'$, this mode has the largest branching ratio if
\begin{equation}
\lambda_u' \gtrsim \lambda_u^2 / y_c \,.
\end{equation}
If this condition is fulfilled, this will be by far the dominant decay mode
and $c'$ pair production will give rise to six jets final states 
\be
pp\to c'c' \to S_c S_c cc \to 6c.
\ee
Presently several searches at high jet multiplicities inspired by supersymmetry, that do not rely on missing $E_T$ cuts, are ongoing at both CMS and ATLAS.

\subsection{Spin 0}

The flavon $S_c$ itself can in principle be detectable at the LHC. Its decay depends on its mass relative to that of the heavy fermion $c'$. Since the coupling to the charm quark is suppressed by the small mixing angles $s_{L,R}^c$, its dominant decay modes will be (depending on the hierarchy between $M_{S_c}$ and $m_{c'}$)
\begin{align}
S_c &\to c'c'\to VVcc,Vhcc,hhcc \, ,
\\
S_c &\to c'c\to Vcc,hcc \, , 
\\
S_c &\to cc \, ,
\end{align}
where $V=W,Z$.

In the last case, $S_c$ could show up as a narrow peak in the dijet invariant mass, since its width is very small compared to its mass
\be
\frac{m_S}{\Gamma_{S_c}^{\mathrm{tot}}}= \frac{8\pi}{(\lambda_u's_L^cs_R^c)^2} \times \text{BR}(S_c\to \bar cc)
\approx
\frac{8\pi m_c^2}{\langle S_c \rangle^2}
\,.
\ee
However, the production cross section is tiny,
\be
\sigma(pp\to~ S_c) = (\lambda_u' s_L^c s_R^c)^2\frac{ \pi}{2s}
 \times \mathcal L_{c\bar c}\!\left(\tfrac{m_{S_c}^2}{s}\right)
\approx 
\frac{\pi m_c^2}{s\langle S_c \rangle^2}
 \times \mathcal L_{c\bar c}\!\left(\tfrac{m_{S_c}^2}{s}\right)
\ee
where $s$ is the invariant mass squared in the $pp$ rest frame and $\mathcal L_{c\bar c}(\hat s)$ the $c\bar c$ parton-parton luminosity function evaluated at the partonic invariant mass squared $\hat s$.

\begin{figure}[tbp]
\label{fig: S_xs}
 \centering%
\includegraphics[width=0.6\textwidth]{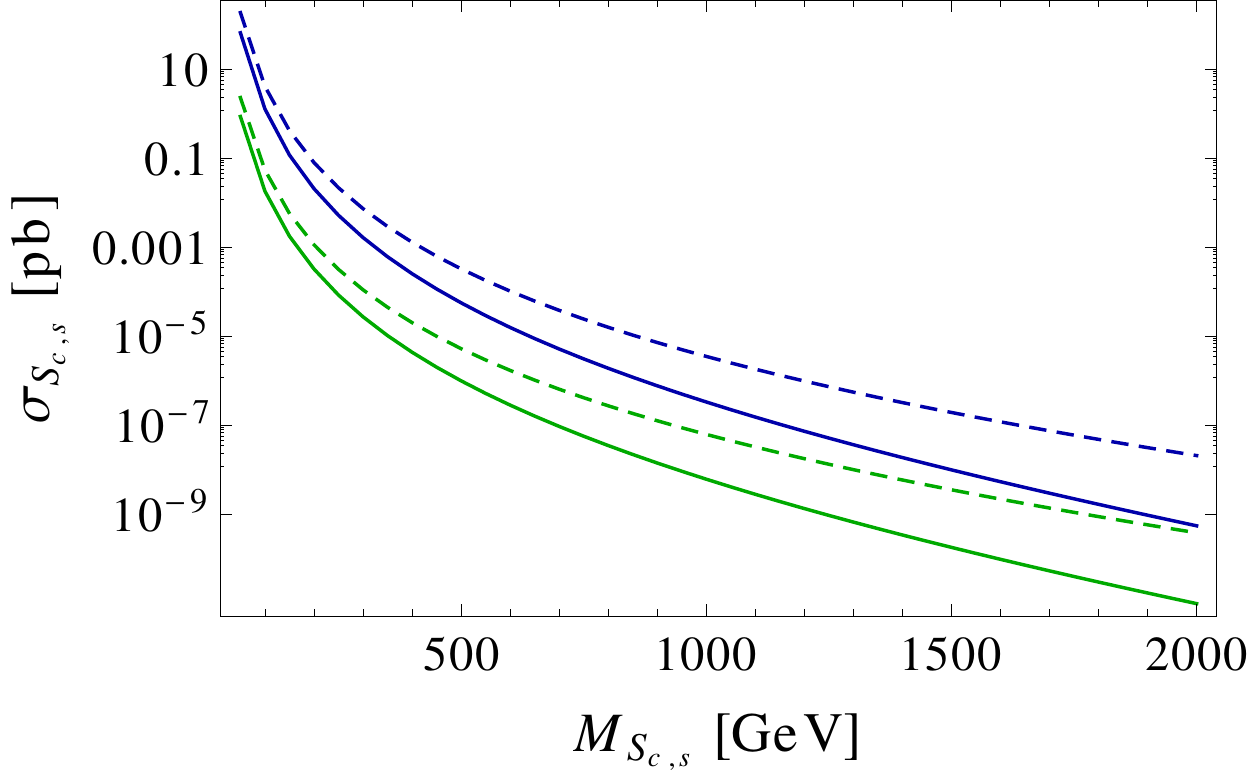}
 \caption{The production cross sections of the lightest flavons $S_{c}$ (in green) and $S_{s}$ (in blue) at the LHC with $\sqrt{s}=8$~TeV (solid) or 14~TeV (dashed) for different values of their mass.}
\end{figure}

The numerical production cross sections for $S_c$ and $S_s$ at LHC for $\sqrt{s}= 7$ and 14~TeV,
using MSTW 2008 LO PDFs \cite{Martin:2009iq},
are shown in fig.~\ref{fig: S_xs}. These cross sections are below current and future limits \cite{Abazov:2003tj,  Aaltonen:2008dn, :2010bc, Khachatryan:2010jd, Aad:2011fq, Chatrchyan:2011ns}.
For example, we find that the $S_c$ production cross section at $\sqrt{s}= 7$~TeV is more than 5 orders of magnitude smaller than the current dijet bound from ATLAS at $M_{S_c}=800$~GeV and more than 7 orders of magnitude smaller at $M_{S_c}=1.5$~TeV. The Tevatron bounds at lower masses are evaded just as easily. It is appropriate to say, however, that most of the limits from searches on the dijet invariant mass spectrum are model dependent or obtained postulating a gaussian shape for the resonance. Here we have quoted the ATLAS limit on axigluons and excited quarks and it is not unreasonable to expect $O(1)$ differences in the bound on our specific model, mainly due to initial and final state radiation. Nonetheless it is clear that our statements remain correct and that we can safely neglect also pair production of the scalar flavons.

\section{Summary}

We have studied the phenomenology of a model where an $SU(2)^3$ subgroup of the $U(3)^3$ flavour group present in the SM quark sector is gauged, in the spirit of \cite{Grinstein:2010ve}. We have been motivated by the approximate $U(2)^3$ symmetry manifest in the quark masses and mixings. Our main findings can be summarized as follows.
\begin{enumerate}
\item Since the third generation quarks are singlets under the flavour gauge group, observables related to the third generation, in particular electroweak precision and $B$ physics observables, are virtually unaffected.
\item\label{it:V} The strongest constraint in the flavour sector is given by $|\epsilon_K|$. Tree-level contributions mediated by flavour gauge bosons (FGBs) lead to a lower bound on the 3 FGBs of $SU(2)_Q$ of about $10^8$~GeV, which affects also the order of magnitude of third generation fermion partner masses.
\item First and second generation fermion partner masses are weakly constrained by flavour physics. The strongest constraint comes from $W$-$c'$ box contributions to neutral kaon mixing, which lead to an enhancement of $|\epsilon_K|$ that are generically of phenomenologically viable size.
\item The second generation fermion partners $s'$ and $c'$ can be light and might be produced at the LHC. The resulting signatures depend on the mass of the scalar flavons $S_s$ and $S_c$, that are expected to be comparable to  $s'$ and $c'$.
\begin{itemize}
\renewcommand{\labelitemi}{$-$}
\item If the flavons are heavier than the fermions, direct bounds from Tevatron and the LHC imply $m_{c',s'}\gtrsim300$~GeV and the pair-produced fermions will each decay to $Wq$, $hq$ or $Zq$ with a ratio $2:2:1$.
\item If the flavons are lighter than the fermions, $c'$ ($s'$) pair production will lead to six-jet final states, which are not strongly constrained yet.
\end{itemize}
\item Direct production of the scalars is suppressed to a level completely negligible with respect to the backgrounds.
\end{enumerate}

While point \ref{it:V}.\ above is clearly a strong constraint, we find interesting that with respect to the original model \cite{Grinstein:2010ve} and its phenomenology \cite{Grinstein:2010ve,Buras:2011wi}, several serious constraints are evaded -- including electroweak precision observables,  $V_{tb}$, and $\Delta M_{B_{d,s}}$. In the $SU(2)^3$ case box contributions to kaon mixing are well in agreement with the data and the collider phenomenology is quite different. Finally, the smaller number of parameters makes the model even more predictive than in the $SU(3)^3$ case, with the fixed branching ratios of the $c'$ in eq.~(\ref{eq:BRSM}) being one example.
Consequently, the model has good prospects of being tested at the LHC if the second generation partners are light enough.

\section*{Acknowledgments}
We thank Riccardo Barbieri for useful discussions and for a careful reading of the manuscript.
This work was supported in part by the EU ITN ``Unification in the LHC Era'', 
contract PITN-GA-2009-237920 (UNILHC).

\appendix

\section{Fermion mass matrix diagonalization}\label{sec:fermiondiag}

Grouping the fermion fields into six-dimensional vectors of left- and right-handed up- and down-type fields, e.g. $\tilde{f}_{uL} = (U_L^1~U_L^2~T_L~\psi_u^1~\psi_u^2~\psi_t)^T$, the fermion mass terms can be diagonalized by 
unitary field rotations, $\tilde{f}_{uL} = \mathcal U_{uL}\, f_{uL}$ etc., where the fields on the right-hand side are in the mass eigenbasis.
The fermion mass matrix can be diagonalized for up-type quarks (and analogously for down-type quarks) as
\begin{equation}
M_u^\text{diag} =
\mathcal U_{uL}^\dagger\,
\left(
\begin{array}{cccccc}
 0 & 0 & 0 & c_u v & 0 & 0 \\
 0 & 0 & 0 & 0 & c_u v & \lambda_q V \\
 0 & 0 & v y_t & 0 & 0 & 0 \\
 M_u & 0 & 0 & \frac{c_u c_u M_u}{y_u} & \frac{e^{-i \alpha_u}
   c_u M_u s_u}{y_u} & 0 \\
 0 & M_u & 0 & -\frac{e^{i \alpha_u} c_u M_u s_u}{y_c} &
   \frac{c_u c_u M_u}{y_c} & 0 \\
 0 & 0 & \lambda_t v & 0 & 0 & M_q'
\end{array}
\right)
\,\mathcal U_{uR}
\,.
\end{equation}
The couplings to the $Z$ in the mass eigenbasis (using Dirac spinor notation) are
\begin{equation}
\mathcal L \supset \frac{g}{c_w} \bar f_{u} \gamma_\mu \left(
\tfrac{1}{2}\;\mathcal U_{uL}^\dagger\,D_L\,\mathcal U_{uL}P_L
+
\tfrac{1}{2}\;\mathcal U_{uR}^\dagger\,D_R\,\mathcal U_{uR}P_R
-\tfrac{2}{3}s_w^2
\right)
f_{u} \,Z^\mu
\,,
\end{equation}
where $D_L=\text{diag}(1,1,1,0,0,1)$ and $D_R=\text{diag}(0,0,0,0,0,1)$, and the couplings to the $W$ read
\begin{equation}
\mathcal L \supset \frac{g}{\sqrt{2}} \bar f_{u} \gamma_\mu \left(
\mathcal U_{uL}^\dagger\,D_L\,\mathcal U_{dL}P_L
+
\mathcal U_{uR}^\dagger\,D_R\,\mathcal U_{dR}P_R
\right)
f_{d} \,W^{+\mu}
\,.
\end{equation}
The non-zero right-handed $Z$ and $W$ couplings induced by the couplings of the field $\psi_{q_R}$ are phenomenologically irrelevant given the large mass and tiny mixings for the $t'$ and $b'$ we found in our analysis.
Neglecting them and decomposing the $6\times6$ unitary mixing matrices $\mathcal U_{qA}$ into (non-unitary) $3\times3$ submatrices as
\begin{equation}
\mathcal U_{qA} =
\begin{pmatrix}
U_{qA}^l & U_{qA}^{lh} \\
U_{qA}^{hl} & U_{qA}^h
\end{pmatrix},
\end{equation}
the $W$ couplings have the same form as in the SM with the CKM matrix replaced by the non-unitary matrix
\begin{equation}
V=U_{uL}^{l\dagger}U_{dL}^l
\,,
\end{equation}
which is to a good approximation the same as (\ref{eq:ckm}).

\bibliographystyle{My}
\bibliography{su2gauge}

\end{document}